\documentclass[a4paper, 12pt]{article}
\usepackage{jheppub}
\usepackage{amssymb} 
\usepackage{amsmath}
\usepackage{mathtools}
\usepackage{amsfonts}    
\usepackage{dsfont}
\usepackage{young}
\usepackage[vcentermath]{youngtab}
\usepackage[mathscr]{euscript}
\usepackage{tensor}
\usepackage{xcolor}
\usepackage{braket}

\newif\ifdetails
\detailstrue

\allowdisplaybreaks

\title{\boldmath Tensionless hybrid strings in $\rm AdS_3\times S^3\times S^3\times S^1$: Free field realisation}

\author[]{Vit Sriprachyakul}
\affiliation[]{Department of Particle Physics and Astrophysics, Weizmann Institute of Science,\\ Rehovot 7610001, Israel}
\emailAdd{vit.sriprachyakul@weizmann.ac.il}

\abstract{We discuss a Wakimoto-like free field realisation of ${\frak d}(2,1;\alpha)_1$, whose $\frak{sl}(2,\mathbb{R})$ subalgebra has level $k=1$, that requires no gauging, i.e., realises the current algebra exactly. We then compute the partition function of the theory and show that, by combining this with the ghost contribution, the full, on-shell projected string partition function reproduces precisely the single-particle partition function of the ${\rm Sym}^N({\cal S'}_0^2)$ theory, i.e. the symmetric orbifold theory of 8 free fermions, 1 compact free boson, and 1 non-compact free boson. We also discuss other aspects such as DDF operators and BRST and physical state conditions.}

\begin{document}

\maketitle

\section{Introduction}

29 years have passed since the appearance of the celebrated paper \cite{Maldacena:1997re} that started the AdS/CFT programme. Nonetheless, it remains a rather active field of research with still a lot of things to be explored and understood. Of all the family of ${\rm AdS}_{d+1}$ backgrounds, $\rm AdS_3$ stood out as one of the best-suited backgrounds to be studied from the worldsheet point of view \cite{Maldacena:2000hw, Maldacena:2000kv, Maldacena:2001km, Giveon:1998ns, Kutasov:1999xu, Elitzur:1998mm,Eberhardt:2017fsi}: mainly because the purely-NSNS supported $\rm AdS_3$ geometry can be described in terms of the conventional RNS worldsheet theory. Furthermore, inside the family of $\rm AdS_3$ spacetimes parametrised by the amount of (quantised) NSNS flux $k$, the tensionless limit ($k=1$) has received attention recently, and the amount of work is accumulating \cite{Gaberdiel:2018rqv, Eberhardt:2018ouy, Eberhardt:2019ywk, Eberhardt:2025sbi, Dei:2020zui, Dei:2023ivl, Dei:2024sct}. The classic example of the tensionless ${\rm AdS}_3\times X$ background is when $X={\rm S^3}\times\mathbb{T}^4$, whose dual is the symmetric orbifold of $\mathbb{T}^4$. Despite the obstacle that the string theory can only be formulated in the hybrid formalism \cite{Berkovits:1999im}, due to the non-unitary problem in the $\rm S^3$ theory,\footnote{This is because the $\rm S^3$ theory is given by the $\frak{su}(2)^{(1)}_1$ WZW theory. After decoupling the fermions, we obtain $\frak{su}(2)^{(1)}_1=\frak{su}(2)_{-1}+\text{3 free fermions}$. The negative level of the bosonic algebra implies that $c_{\frak{su}(2)_{-1}}=\frac{3(-1)}{(-1)+2}=-3<0$ whereas unitary CFTs would satisfy $c\geq0$.} many aspects of the equivalence have been studied: spectrum \cite{Gaberdiel:2018rqv, Eberhardt:2018ouy}, correlation functions \cite{Eberhardt:2019ywk, Dei:2020zui, Dei:2023ivl}, D-branes \cite{Gaberdiel:2021kkp}, topological defects \cite{Knighton:2024noc}, and $T\bar T$ deformations \cite{Dei:2024sct} for instance. On the contrary, despite being able to be described in the conventional RNS language, the tensionless background $\rm AdS_3\times S^3\times S^3\times S^1$ is relatively unexplored, see recent developments \cite{Gaberdiel:2018rqv,Giribet:2018ada,Gaberdiel:2024dva, Belleri:2025eun}. In this article, we take a slightly unconventional turn and study the tensionless background $\rm AdS_3\times S^3\times S^3\times S^1$ in the hybrid formalism \cite{Berkovits:1999im, Eberhardt:2019niq}.

The hybrid formulation of strings in $\rm AdS_3\times S^3\times S^3\times S^1$ involves the following field content on the worldsheet \cite{Eberhardt:2019niq}:
\begin{itemize}

\item The current algebra ${\frak d}(2,1;\alpha)_k$, where $k$ denotes the level of the ${\frak {sl}}(2,\mathbb{R})$ subalgebra.

\item A compact (bosonic) ${\frak u}(1)$ theory.

\item The usual conformal $bc$ ghosts and a chiral boson $\rho$ with background charge $3$ (this gives central charge $c=1+3(3)^2=28$), which is the usual hybrid ghost. It is also standard in the hybrid language to bosonise the conformal $bc$ ghosts.

\item Two $bc$ systems of weights $(1,0)$ which we will denote by $(b',c')$ and $(b'',c'')$.

\end{itemize}
Apart from the current algebra, the remaining ingredients can be described in terms of free fields. Thus, one of the main challenges in applying the hybrid formalism in the background $\rm AdS_3\times S^3\times S^3\times S^1$ lies in finding a nice treatment of the current algebra. In $\rm AdS_3\times S^3\times\mathbb{T}^4$, it was shown that when the level $k$ of $\frak{sl}(2,\mathbb{R})$ in $\frak{psu}(1,1|2)_k$ is 1, the current algebra admits free field realisations: one in terms of symplectic bosons and free fermions \cite{Dei:2020zui} and another in terms of a Wakimoto-like representation \cite{Dei:2023ivl}. The tensionless limit\footnote{Recall that this means 2 units of NSNS fluxes through the 2 $\rm S^3$'s and 1 unit of NSNS flux through $\rm AdS_3$.} of strings in $\rm AdS_3\times S^3\times S^3\times S^1$ was shown to admit a free field realisation in terms of symplectic bosons and free fermions \cite{Gaberdiel:2024dva}. However, given the technology developed in, for instance, correlation computation \cite{Dei:2023ivl, Knighton:2023mhq, Knighton:2024qxd} and OPE analysis \cite{Sriprachyakul:2025ubx}, a Wakimoto-like free field realisation is highly desired. In this article, we construct such a realisation and use it to discuss string partition function, BRST current, and DDF operators.

This article is organised as follows. In Section \ref{sec: hybrid formulation}, we describe our free field realisation, giving ${\frak d}(2,1;\alpha)_1$ currents and stress tensor in terms of the free fields and constructing the Hilbert space, and show that it reproduces the right partition function of the dual CFT, that is, the single-particle partition function of the symmetric orbifold theory of 8 free fermions, 1 compact free boson, and 1 non-compact free boson, denoted by ${\rm Sym}^N({\cal S'}_0^2)$. The prime emphasises that ${\cal S'}^2_0$ contains 1 compact and 1 non-compact free bosons and thus, is not the standard ${\cal S}^2_0$ which contains 2 compact free bosons. We also discuss the physical state condition and the relevant DDF operators, explaining how to construct them in the hybrid language. We then conclude and discuss several future directions in Section \ref{sec: conclusion and outlook}. Appendix \ref{appendix: current algebra convention} summarises our convention for the current algebra and Appendix \ref{appendix: RNS to hybrid} elaborates the translation from the conventional RNS to the hybrid formulations, giving explicit expressions for the DDF operators that are too cumbersome to include in the main body. We also discuss how to construct the worldsheet ${\cal N}=4$ algebra which allows rewriting RNS physical state conditions in terms of ${\cal N}=4$ generators. Appendix \ref{appendix: theta function convention} states our definitions for the theta functions and modular forms.

\section{Hybrid formulation}\label{sec: hybrid formulation}

In this section, we discuss the Wakimoto-like free field description of $\frak{d}(2,1;\alpha)_1$. The convention of the current algebra is spelled out in Appendix \ref{appendix: current algebra convention}. We then use this free field realisation to compute the full string partition function and show that, upon the on-shell projection, it reproduces exactly the expected single-particle partition function of the dual symmetric orbifold theory ${\rm Sym}^N({\cal S'}_0^2)$. We then discuss the BRST and DDF operators.

\subsection{The free field realisation and the partition function}

Consider a $\beta\gamma$ system of weights $(1,0)$, a linear dilaton field $\Phi$ of background charge $\sqrt{2}$, and four systems of $(p^{\alpha\beta},\theta^{\gamma\delta})$\footnote{Intuitively, the $(\gamma,\theta^{\alpha\beta})$ form the ${\cal N}=4$ superspace coordinates in the spacetime, see also the discussion in \cite{Dei:2023ivl}.} with weights $(1,0)$ where the Greek indices run over $\{+,-\}$. These free fields satisfy the OPEs
\begin{equation}
\beta(z)\gamma(w)\sim-\frac{1}{z-w},\quad \Phi(z)\Phi(w)\sim-\ln|z-w|^2,\quad p^{\alpha\beta}(z)\theta^{\gamma\delta}(w)\sim\frac{\epsilon^{\alpha\gamma}\epsilon^{\beta\delta}}{z-w}\,,
\end{equation}
where $\epsilon^{+-}=-\epsilon^{-+}=1$. As a sanity check, the total central charge is $2+(1+3(\sqrt{2})^2)+4(-2)=1$ which is precisely the central charge of $\frak{d}(2,1;\alpha)_k$ for any $k$. Indeed, one can directly verify that the generators of $\frak{d}(2,1;\alpha)_1$ are given by
\begin{equation}
\begin{gathered}
J^+=\beta,\quad J^3=\frac{1}{\sqrt{2}}\partial\Phi+\beta\gamma+\frac{1}{2}\epsilon_{\alpha\gamma}\epsilon_{\beta\delta}p^{\alpha\beta}\theta^{\gamma\delta},\\
J^-=\beta\gamma^2-\partial\gamma+\sqrt{2}\gamma\partial\Phi+\epsilon_{\alpha\gamma}\epsilon_{\beta\delta}p^{\alpha\beta}\theta^{\gamma\delta}\gamma,\\
K^{(+)a}=(\sigma^a)_{\alpha\gamma}\epsilon_{\beta\delta}p^{\alpha\beta}\theta^{\gamma\delta},\quad K^{(-)a}=\epsilon_{\alpha\gamma}(\sigma^a)_{\beta\delta}p^{\alpha\beta}\theta^{\gamma\delta}\,,
\label{eq: bosonic d(2,1) currents}
\end{gathered}
\end{equation}
for the bosonic ones and
\begin{equation}
\begin{gathered}
S^{+\alpha\beta}=p^{\alpha\beta}-\beta\theta^{\alpha\beta},\quad S^{-\alpha\beta}=\sqrt{2}\theta^{\alpha\beta}\partial\Phi+\theta^{\alpha\beta}\beta\gamma-p^{\alpha\beta}\gamma-\partial\theta^{\alpha\beta}+\theta^{\alpha\beta}\epsilon_{\mu\rho}\epsilon_{\nu\sigma}p^{\mu\nu}\theta^{\rho\sigma}\,,
\label{eq: fermionic d(2,1) currents}
\end{gathered}
\end{equation}
for the fermionic ones.
Here $\epsilon_{\alpha\beta}$ is the inverse of $\epsilon^{\alpha\beta}$ and the Pauli matrices are given by
\begin{equation}
\begin{gathered}
(\sigma^-)\indices{^\alpha_\beta}=
\begin{pmatrix}
0 & -1\\
0 & 0
\end{pmatrix},\quad
(\sigma^3)\indices{^\alpha_\beta}=\frac{1}{2}
\begin{pmatrix}
-1 & 0\\
0 & 1
\end{pmatrix},\quad
(\sigma^+)\indices{^\alpha_\beta}=
\begin{pmatrix}
0 & 0\\
-1 & 0
\end{pmatrix}\,,\\
(\tau^-)\indices{^\alpha_\beta}=
\begin{pmatrix}
0 & 1\\
0 & 0
\end{pmatrix},\quad
(\tau^3)\indices{^\alpha_\beta}=\frac{1}{2}
\begin{pmatrix}
-1 & 0\\
0 & 1
\end{pmatrix},\quad
(\tau^+)\indices{^\alpha_\beta}=
\begin{pmatrix}
0 & 0\\
-1 & 0
\end{pmatrix}\,,
\label{eq: Pauli matrices convention}
\end{gathered}
\end{equation}
where we have simultaneously defined the `$\tau$' matrices which are for later convenience. The Greek indices are raised and lowered by the Levi-Civita symbol $\epsilon^{\alpha\beta}$ while the Latin (adjoint) indices $a$ are raised and lowered by $h^{ab}_{\frak{su}}$ for the Pauli matrices and by $h^{ab}_{\frak{sl}}$ for the $\tau$ matrices. Explicitly, the Killing forms read
\begin{equation}
\begin{gathered}
h^{ab}_{\frak{sl}}=
\begin{pmatrix}
0 & 2 & 0\\
2 & 0 & 0\\
0 & 0 & -1
\end{pmatrix}=-2{\rm Tr}(\tau^a\tau^b),\quad h^{ab}_{\frak{su}}=
\begin{pmatrix}
0 & 2 & 0\\
2 & 0 & 0\\
0 & 0 & 1
\end{pmatrix}=2{\rm Tr}(\sigma^a\sigma^b)\,.
\label{eq: Killing forms}
\end{gathered}
\end{equation}
The normal-ordering convention here is as in \cite{Polchinski:1998rq} (see also Appendix C of \cite{Gaberdiel:2022als} for a quick review and some useful formulas). The stress tensor is given by
\begin{equation}
\begin{split}
T=-\beta\partial\gamma-\epsilon_{\alpha\gamma}\epsilon_{\beta\delta}p^{\alpha\beta}\partial\theta^{\gamma\delta}-\frac{1}{2}(\partial\Phi)^2-\frac{1}{\sqrt{2}}\partial^2\Phi\,.
\end{split}
\end{equation}
Using this stress tensor, one can easily verify that all the currents have weight 1 and are primary.

\subsubsection{Representations and Hilbert space}
In order to write down the partition function, we first define our Hilbert space. Since we have only free fields, it is rather straightforward to define the vacuum state. We will define the ground state $\ket{m,j}$ to satisfy the following
\begin{equation}
\begin{gathered}
\beta_0\ket{m,j}=(m+j)\ket{m+1,j},\quad \gamma_0\ket{m,j}=\ket{m-1,j},\\
p^{\alpha\beta}_0\ket{m,j}=0,\quad (\partial\Phi)_0\ket{m,j}=-\sqrt{2}j\ket{m,j}\,,
\end{gathered}
\end{equation}
where $m\in\mathbb{Z}+\alpha$, $\alpha\in[0,1)$ and $j=-\tfrac{1}{2}+is,s\in\mathbb{R}$.\footnote{This particular value of $j$ is so that the states form the continuous representions ${\cal C}^j_{\alpha}$ of $\frak{d}(2,1;\alpha)_1$, see also \cite{Gaberdiel:2024dva}.}
The other half of the fermion zero modes $\theta_0^{\alpha\beta}$ thus generate the states
\begin{equation}
\begin{gathered}
\ket{m,j},\quad \theta^{++}_0\ket{m,j},\quad \theta^{+-}_0\ket{m,j},\quad \theta^{-+}_0\ket{m,j},\quad \theta^{--}_0\ket{m,j},\quad \theta^{++}_0\theta^{+-}_0\ket{m,j},\\
\theta^{++}_0\theta^{-+}_0\ket{m,j},\quad
\theta^{++}_0\theta^{--}_0\ket{m,j},\quad \theta^{+-}_0\theta^{-+}_0\ket{m,j},\quad \theta^{+-}_0\theta^{--}_0\ket{m,j},\quad \theta^{-+}_0\theta^{--}_0\ket{m,j},\\
\theta^{++}_0\theta^{+-}_0\theta^{-+}_0\ket{m,j}, \quad
\theta^{++}_0\theta^{+-}_0\theta^{--}_0\ket{m,j},\quad \theta^{++}_0\theta^{-+}_0\theta^{--}_0\ket{m,j},\quad \theta^{+-}_0\theta^{-+}_0\theta^{--}_0\ket{m,j},\\
\theta^{++}_0\theta^{+-}_0\theta^{-+}_0\theta^{--}_0\ket{m,j}\,.
\end{gathered}
\end{equation}
These can be rearranged into the $\bf(m,n)$-dimensional representations of the zero modes of $\mathfrak{su}(2)_2\otimes\mathfrak{su}(2)_2$ as follows
\begin{equation}
\begin{split}
{\bf(1,1)}=&\{\ket{m,j}\},\quad \{\theta^{++}_0\theta^{+-}_0\theta^{-+}_0\theta^{--}_0\ket{m,j}\}\,,\\
{\bf(2,2)}=&\{\theta^{++}_0\ket{m,j}, \theta^{+-}_0\ket{m,j}, \theta^{-+}_0\ket{m,j}, \theta^{--}_0\ket{m,j}\}\,,\\
&\{\theta^{++}_0\theta^{+-}_0\theta^{-+}_0\ket{m,j},
\theta^{++}_0\theta^{+-}_0\theta^{--}_0\ket{m,j}, \theta^{++}_0\theta^{-+}_0\theta^{--}_0\ket{m,j}, \theta^{+-}_0\theta^{-+}_0\theta^{--}_0\ket{m,j}\}\,,\\
{\bf (3,1)}=&\{\theta^{++}_0\theta^{+-}_0\ket{m,j},(\theta^{-+}_0\theta^{+-}_0+\theta^{++}_0\theta^{--}_0)\ket{m,j},\theta^{-+}_0\theta^{--}_0)\ket{m,j}\}\,,\\
{\bf (1,3)}=&\{\theta^{++}_0\theta^{-+}_0\ket{m,j},(\theta^{-+}_0\theta^{+-}_0-\theta^{++}_0\theta^{--}_0)\ket{m,j},\theta^{+-}_0\theta^{--}_0\ket{m,j}\}\,,
\end{split}
\end{equation}
where we note that $K^{(\pm)a}_0\ket{m,j}=0$ from the definition of the ground state $\ket{m,j}$. To determine the $\mathfrak{sl}(2,\mathbb{R})$ spin $j'$, we check the action of the zero modes of the $\mathfrak{sl}(2,\mathbb{R})$ currents and we have
\begin{equation}
\begin{gathered}
J^+_0\ket{m,j}=(m+j)\ket{m+1,j},\quad J^3_0\ket{m,j}=m\ket{m,j},\\
J^-_0\ket{m,j}=(m-j)\ket{m-1,j}\,.
\label{eq: sl(2,R) zero modes action}
\end{gathered}
\end{equation}
The $\mathfrak{sl}(2,\mathbb{R})$ quadratic Casimir is given by
\begin{equation}
{\cal C}_{\mathfrak{sl}}=\frac{1}{2}(J^+_0J^-_0+J^-_0J^+_0)-J^3_0J^3_0=j'(1-j')\,,
\end{equation}
and from \eqref{eq: sl(2,R) zero modes action}, we obtain
\begin{equation}
{\cal C}_{\mathfrak{sl}}\ket{m,j}=j(1-j)\,.
\end{equation}
Hence, we see that the $\mathfrak{sl}(2,\mathbb{R})$ spin of $\ket{m,j}$ is
\begin{equation}
j'=j\,.
\end{equation}
Next, we note that
\begin{equation}
\begin{gathered}
[J^+_0,\theta^{\alpha\beta}_0]=0,\quad [J^3_0,\theta^{\alpha\beta}_0]=-\frac{1}{2}\theta^{\alpha\beta}_0,\quad [J^-_0,\theta^{\alpha\beta}_0]=-(\theta^{\alpha\beta}\gamma)_0\,.
\end{gathered}
\end{equation}
Using this, we conclude that\footnote{Below, we use a shorthand where the exponent of $\theta^{\alpha\beta}$ denotes how many zero modes of $\theta^{\alpha\beta}_0$ are acting on the ground state $\ket{m,j}$. For example, the states $\theta^{++}_0\theta^{--}_0\ket{m,j}$ and $\theta^{+-}_0\theta^{--}_0\ket{m,j}$ both belong to the class $(\theta^{\alpha\beta})^2\ket{m,j}$.}
\begin{equation}
\begin{split}
j'\Bigl(\theta^{\alpha\beta}\ket{m,j}\Bigr)=&j+\frac{1}{2},\\
j'\Bigl((\theta^{\alpha\beta})^2\ket{m,j}\Bigr)=&j+1,\\
j'\Bigl((\theta^{\alpha\beta})^3\ket{m,j}\Bigr)=&j+\frac{3}{2},\\
j'\Bigl((\theta^{\alpha\beta})^4\ket{m,j}\Bigr)=&j+2,\\
\end{split}
\end{equation}
or equivalently, the representation diamond is
\begin{equation}
\begin{gathered}
({\cal C}^j_\alpha,{\bf1,1})\\
({\cal C}^{j+\frac{1}{2}}_{\alpha-\frac{1}{2}},{\bf2,2})\\
({\cal C}^{j+1}_\alpha,{\bf3,1}),\quad ({\cal C}^{j+1}_\alpha,{\bf1,3})\\
({\cal C}^{j+\frac{3}{2}}_{\alpha-\frac{1}{2}},{\bf2,2})\\
({\cal C}^{j+2}_\alpha,{\bf1,1})\,.
\label{eq: long representation diamond}
\end{gathered}
\end{equation}
We define the spectral flow actions as follows
\begin{equation}
\begin{aligned}
\sigma^w(\beta_n)=&\beta_{n-w},&\quad \sigma^w(\gamma_n)=&\gamma_{n+w},\quad \sigma^w((\partial\Phi)_n)=&(\partial\Phi)_n+\frac{w}{\sqrt{2}}\delta_{n,0},\\
\sigma^w(p^{\alpha\beta}_n)=&p^{\alpha\beta}_{n+w\frac{\alpha-1}{2}},&\quad \sigma^w(\theta^{\alpha\beta}_n)=&\theta^{\alpha\beta}_{n+w\frac{\alpha+1}{2}}\,.
\label{eq: spectral flow actions on the free fields}
\end{aligned}
\end{equation}
This definition implies the spectral flow actions
\begin{equation}
\begin{aligned}
\sigma^w(K^{(+)3}_n) & =  K^{(+)3}_n+w\delta_{n,0}\ ,\qquad & \sigma^w(K^{(+)\pm}_n) & =  K^{(+)\pm}_{n\pm w}\ , \\
\sigma^w(J^{3}_n) & = J^{3}_n+\frac{w}{2}\delta_{n,0}\ ,\qquad & \sigma^w(J^{\pm}_n) & =  J^{\pm}_{n\mp w}\ , \\
\sigma^w(K^{(-)a}_n) & =  K^{(-)a}_n \ , \qquad & \sigma^w(L_0) & =  L_0+w(K^{(+)3}_0-J^3_0)+\frac{w^2}{4}\,.
\label{eq: spectral flow actions on the currents}
\end{aligned}
\end{equation}
The Hilbert space is taken to be the diagonal combination, that is,
\begin{equation}
{\cal H} =  \left[ \sum_{w\in\mathbb{Z}} \int_0^1d \alpha \int_{\mathds{R}} ds\, \sigma^w \big( {\cal C}^j_\alpha \bigr) \otimes \sigma^w \bigl( \overline{{\cal C}^j_\alpha} \bigr) \right] \otimes {\cal H}^{S^1}\,,
\label{eq: diagonal Hilbert space}
\end{equation}
where we note that ${\cal C}^j_\alpha$ denotes the \emph{whole} diamond \eqref{eq: long representation diamond}.
Next, we want to write down the full string partition function. Firstly, the (unflowed) character we are interested in is 
\begin{equation}
{\rm Tr}\left( q^{L_0-\frac{c}{24}}x^{J^3_0}y_+^{K^{(+)3}_0}y_-^{K^{(-)3}_0} \right)\,,
\end{equation}
where $q=\exp(2\pi i\tau)$, $x=\exp(2\pi it)$, $y_\pm=\exp(2\pi iz_\pm)$. Since each free field theory decouples from one another, we can calculate the character associated to each theory separately. The $\beta\gamma$ character reads
\begin{equation}
\sum_{m\in\mathbb{Z}+\alpha}\frac{x^{m+j}}{\eta^2}\,,
\end{equation}
whereas the fermion character reads
\begin{equation}
\frac{x\vartheta_2(\frac{t+z_++z_-}{2};\tau)\vartheta_2(\frac{t+z_+-z_-}{2};\tau)\vartheta_2(\frac{t-z_++z_-}{2};\tau)\vartheta_2(\frac{t-z_+-z_-}{2};\tau)}{\eta^4}\,,
\end{equation}
and lastly the linear dilaton character is (recall that $j=-\frac{1}{2}+is$)\footnote{From the stress tensor, we have that $L_0-\tfrac{c}{24}=-j(1+j)-\frac{c}{24}=\tfrac{1}{4}+s^2-\frac{7}{24}=s^2-\frac{1}{24}$.}
\begin{equation}
\frac{q^{s^2}x^{-j}}{\eta}\,.
\end{equation}
Combining these pieces gives exactly the result obtained in \cite{Gaberdiel:2024dva}
\begin{equation}
\begin{split}
\chi_{{\cal C}^j_\alpha}(\tau;z_\pm,t)=q^{s^2}\sum_{r\in\mathbb{Z}+\alpha}x^r\frac{\vartheta_2(\frac{t+z_++z_-}{2};\tau)\vartheta_2(\frac{t+z_+-z_-}{2};\tau)\vartheta_2(\frac{t-z_++z_-}{2};\tau)\vartheta_2(\frac{t-z_+-z_-}{2};\tau)}{\eta^7}\,.
\label{eq: unflowed free field character}
\end{split}
\end{equation}
Finding the flowed characters proceeds as in \cite{Gaberdiel:2024dva}. 
The flowed character thus reads
\begin{equation}
\begin{split}
&\chi_{\sigma^w({\cal C}^j_\alpha)}(z_\pm,t;\tau)\\
&={\rm Tr}_{\sigma^w({\cal C}^j_\alpha)}\left( q^{L_0-\frac{c}{24}}x^{J^3_0}y_+^{K^{(+)3}_0}y_-^{K^{(-)3}_0} \right)\\
&=q^{\frac{w^2}{4}}x^{\frac{w}{2}}y_+^{w}{\rm Tr}_{{\cal C}^j_\alpha}\left( q^{L_0-\frac{c}{24}}(xq^{-w})^{J^3_0}(y_+q^{w})^{K^{(+)3}_0}y_-^{K^{(-)3}_0} \right)\\
&=q^{s^2-\frac{3w^2}{4}}x^{\frac{3w}{2}}\sum_{r\in\mathbb{Z}+\alpha}x^rq^{-rw}\frac{\vartheta_2(\frac{t+z_++z_-}{2};\tau)\vartheta_2(\frac{t+z_+-z_-}{2};\tau)\vartheta_2(\frac{t-z_++z_-}{2};\tau)\vartheta_2(\frac{t-z_+-z_-}{2};\tau)}{\eta^7}\,.
\label{eq: flowed free field character}
\end{split}
\end{equation}
Establishing the modular invariance follows exactly the same steps as in \cite{Gaberdiel:2024dva}, thus, we omit the details here.

The full string partition function is obtained by combining the $\frak{d}(2,1;\alpha)_1$ partition function with the $\rm S^1$ matter and the ghost $\rho,\sigma,b',c',b'',c''$ contributions which read (see, for example, \cite{Eberhardt:2018ouy,Dei:2023ivl})
\begin{equation}
\begin{split}
Z_\sigma=&|\eta^2|^2\,,\\
Z_\rho=&\left|\frac{\eta^2}{\vartheta_2(0;\tau)^2}\right|^2\,,\\
Z_{b'c'}Z_{b''c''}=&\left| \frac{\vartheta_2(0;\tau)^2}{\eta^2}\right|^2=Z_\rho^{-1}\,,
\end{split}
\end{equation}
while the matter partition function is
\begin{equation}
Z_{\rm matter}=Z_{\frak{d}(2,1;\alpha)_1}Z_{\rm S^1}=\sum_{w\in\mathbb{Z}}\int_\mathbb{R}ds\int_0^1d\alpha\left| \chi_{\sigma^w({\cal C}^j_\alpha)}(z_\pm,t;\tau) \right|^2\times Z_{\rm S^1}(\tau)\,.
\end{equation}
Alternatively, the matter piece can be written as
\begin{equation}
\begin{split}
Z_{\rm matter}=&Z_{\rm S^1}\left| \frac{\vartheta_2(\frac{t+z_++z_-}{2};\tau)\vartheta_2(\frac{t+z_+-z_-}{2};\tau)\vartheta_2(\frac{t-z_++z_-}{2};\tau)\vartheta_2(\frac{t-z_+-z_-}{2};\tau)}{\eta^7} \right|^2\\
&\times \sum_{w\in\mathbb{Z}}\int_\mathbb{R}ds\int_0^1d\alpha \left(q^{s^2+\frac{3w^2}{4}}\bar{q}^{s^2+\frac{3w^2}{4}}\sum_{n,\bar n\in\mathbb{Z}}e^{2\pi i(n-\bar n)(\alpha+\frac{3w}{2})}\delta^2(t-w\tau-n)\right)\,,\\
=&Z_{\rm S^1}\left| \frac{\vartheta_2(\frac{t+z_++z_-}{2};\tau)\vartheta_2(\frac{t+z_+-z_-}{2};\tau)\vartheta_2(\frac{t-z_++z_-}{2};\tau)\vartheta_2(\frac{t-z_+-z_-}{2};\tau)}{\eta^7} \right|^2\\
&\times \sum_{w\in\mathbb{Z}}\int_\mathbb{R}ds\int_0^1d\alpha \left| q^{s^2-\frac{3w^2}{4}}x^{\frac{3w}{2}}\sum_{r\in\mathbb{Z}+\alpha}x^rq^{-rw} \right|^2\,,
\label{eq: full string partition function}
\end{split}
\end{equation}
where
\begin{equation}
\delta^2(t-w\tau-n):=\delta(t-w\tau-n)\delta(\bar t-w\bar\tau-\bar n)\,.
\end{equation}
Let us now define, for convenience, the partition function of the ${\cal S'}^2_0$ theory
\begin{equation}
\begin{split}
Z_{{\cal S'}^2_0}:=\int_\mathbb{R}ds\frac{q^{s^2}\bar{q}^{s^2}}{|\eta|^2}Z_{\rm S^1}\left| \frac{\vartheta_2(\frac{t+z_++z_-}{2};\tau)\vartheta_2(\frac{t+z_+-z_-}{2};\tau)\vartheta_2(\frac{t-z_++z_-}{2};\tau)\vartheta_2(\frac{t-z_+-z_-}{2};\tau)}{\eta^4} \right|^2\,.
\end{split}
\end{equation}
Recall that the prime means that ${\cal S'}^2_0$ contains 1 compact and 1 non-compact free bosons and thus, is not the standard ${\cal S}^2_0$ which contains 2 compact free bosons.
We see that
\begin{equation}
\begin{split}
Z_{\rm string}=&Z_{\frak{d}(2,1;\alpha)_1}Z_{\rm S^1}Z_{b'c'}Z_{b''c''}Z_\rho Z_\sigma\\
=&\int_0^1d\alpha\sum_{w,r,\bar r}x^{r+\frac{3w}{2}}q^{-rw-\frac{3w^2}{4}}\bar x^{\bar r+\frac{3w}{2}}\bar q^{-\bar rw-\frac{3w^2}{4}}Z_{{\cal S'}^2_0}(t,z_\pm;\tau)\,.
\end{split}
\end{equation}
Imposing the on-shell condition, we obtain
\begin{equation}
-rw-\frac{3w^2}{4}+h_{\rm seed}=0,\quad -\bar rw-\frac{3w^2}{4}+\bar h_{\rm seed}=0\,,
\end{equation}
where we expand, schematically,\footnote{We slightly abuse the notation below. More precisely, the c.c. (complex conjugate) here means the appropriate right-moving contribution and, in particular, does not necessarily mean that the right-moving expression is exactly the complex conjugate of the left-moving one. For example, in $Z_{\rm S^1}$, the right-moving momentum contribution schematically reads $\bar q^{(m/R-wR)^2}$ while the left-moving piece reads $q^{(m/R+wR)^2}$ where $m,w$ are the momentum and winding respectively.}
\begin{equation}
Z_{{\cal S'}^2_0}(t,z_\pm;\tau)=\sum_{a,b,c,h_{\rm seed}} q^{h_{\rm seed}}x^ay_+^by_-^c\times\text{c.c.}\,.
\end{equation}
Solving this yields
\begin{equation}
r=\frac{h_{\rm seed}}{w}-\frac{3w}{4},\quad \bar r=\frac{\bar h_{\rm seed}}{w}-\frac{3w}{4}\,,
\end{equation}
and
\begin{equation}
h_{\rm seed}-\bar h_{\rm seed}\equiv 0 ~{\rm mod}~w\,.
\end{equation}
Plugging this back in, the on-shell projected string partition function reads\footnote{We have now focused on the in-states and thus, restricted the sum to positive spectral flows, see also the discussion in \cite{Maldacena:2001km, Eberhardt:2018ouy}.} 
\begin{equation}
\begin{split}
Z'_{\rm string}=\sum_{w=1}^\infty x^{\frac{3w}{4}}\bar x^{\frac{3w}{4}}Z^{R'}_{{\cal S'}^2_0}\left(t,z_\pm;\frac{t}{w} \right)\,,
\end{split}
\end{equation}
where the prime on $Z_{\rm string}$ emphasises that we have done the on-shell projection and the superscript $R'$ emphasises that we have done the projection $h_{\rm seed}-\bar h_{\rm seed}\equiv 0 ~{\rm mod}~w$ and that the partition is given by R-sector partition function of the ${\cal S'}^2_0$ theory. Using the Jacobi theta function identity
\begin{equation}
\vartheta_2\left( \frac{z\pm t}{2};\frac{t}{w} \right)=x^{-\frac{w}{8}}e^{\mp\pi i\frac{zw}{2}}
\left\{ 
\begin{array}{ll}
\vartheta_2\left( \frac{z}{2};\frac{t}{w} \right),\quad w\text{ even}\\
\vartheta_3\left( \frac{z}{2};\frac{t}{w} \right),\quad w\text{ odd}
\end{array} \right.\,,
\end{equation}
we get
\begin{equation}
\begin{split}
Z'_{\rm string}=\sum_{w\text{ even}}x^{\frac{w}{4}}\bar x^{\frac{w}{4}}Z^{R'}_{{\cal S'}^2_0}\left(0,z_\pm;\frac{t}{w} \right)+\sum_{w\text{ odd}}x^{\frac{w}{4}}\bar x^{\frac{w}{4}}Z^{NS'}_{{\cal S'}^2_0}\left(0,z_\pm;\frac{t}{w} \right)\,,
\label{eq: on-shell projected full string partition function}
\end{split}
\end{equation}
where $Z^{NS'}_{{\cal S'}^2_0}$ is $Z^{R'}_{{\cal S'}^2_0}$ where all $\vartheta_2$ have been replaced by $\vartheta_3$ and it is the NS-sector partition function of the ${\cal S'}^2_0$ theory (with, again, the projection $h-\bar h\in w\mathbb{Z}$). Comparing this to the single-particle partition function of the symmetric orbifold theory of ${\cal S'}^2_0$, say in \cite{Gaberdiel:2018rqv},\footnote{Strictly speaking in equations (3.3\&3.5) in \cite{Gaberdiel:2018rqv}, they do not include the momentum contribution of the bosons, however, it is straightforward to include this contribution back in.} we see that the on-shell projected string theory partition function \eqref{eq: on-shell projected full string partition function} agrees precisely with the symmetric orbifold single-particle partition function.

\subsection{The physical state condition and the DDF operators}
We now turn to discuss the BRST operator and the DDF operators. The resulting expressions of these operators are often significantly more complicated than their RNS version, hence, we will not discuss all their hybrid
expressions in the main body but will give the complicated expressions in Appendix \ref{appendix: RNS to hybrid}. Rather, we will briefly argue in this section how we can construct the BRST operator and a complete set of DDF operators, relegating the details to the appendix. 

The BRST operator (charge) is the contour integral of the BRST current, which reads
\begin{equation}
\begin{aligned}
j_{BRST}=cT^m+\hat\gamma G^m+bc\partial c+\frac{3}{4}\partial c\hat\beta\hat\gamma+\frac{1}{4}c\partial\hat\beta\hat\gamma-\frac{3}{4}c\hat\beta\partial\hat\gamma-b\hat\gamma^2\,
\label{eq: RNS BRST current}
\end{aligned}
\end{equation}
in the RNS formalism. Here, $\hat{\beta},\hat{\gamma}$ are the standard superconformal ghosts in the RNS prescription. Following \cite{Berkovits:1999im,Eberhardt:2019niq}, we regard the topologically twisted fermions $p^{\alpha\beta},\theta^{\gamma\delta}$ as coming from refermionising the RNS fermions and the $\hat\beta,\hat\gamma$ ghosts. Hence, the BRST current in the hybrid formulation is nothing but a rewriting of the expression above in terms of the new hybrid variables, see \eqref{eq: the BRST current in hybrid variables}. In fact, one can modify the BRST current by total derivative terms and promote it to one of the supercurrents of a certain small ${\cal N}=4$ algebra on the worldsheet following \cite{Berkovits:1999im}, we discuss this in a bit more detail in Appendix \ref{appendix: RNS to hybrid}. In addition to the BRST current, there are various DDF operators we can construct in the RNS formalism. For instance, the spacetime stress tensor is given by \cite{Giveon:1998ns} (here, we use the same convention in \cite{Sriprachyakul:2024xih})
\begin{equation}
\begin{split}
{\cal L}_m=&\oint\Bigl( \beta\gamma^{m+1}-\frac{(m+1)\gamma^m\partial\Phi}{\sqrt{2}}+m(m+1)\gamma^{m-1}\psi^3\psi^--(m^2-1)\gamma^m\psi^+\psi^-\\
&\hspace{1cm}+m(m-1)\gamma^{m+1}\psi^+\psi^3 \Bigr)\,.
\end{split}
\end{equation}
Here $\psi^a$ are the usual $\rm AdS_3$ fermions in the RNS prescription. The hybrid expression is slightly more complicated \eqref{eq: spacetime stress tensor in hybrid variables}.
The spacetime R symmetry currents ${\cal K}^{(\pm)a}_m$ can also be constructed straightforwardly following \cite{Giveon:1998ns}. The RNS expressions read
\begin{equation}
\begin{aligned}
{\cal K}^{(\pm)3}_n=&\oint\left( \gamma^n\chi^{(\pm)+}\chi^{(\pm)-}-\sqrt{2}n\gamma^{n-1}\psi^-_\gamma\chi^{(\pm)3} \right)\,,\\
{\cal K}^{(\pm)+}_n=&\oint\left( -2\gamma^n\chi^{(\pm)+}\chi^{(\pm)3}-\sqrt{2}n\gamma^{n-1}\psi^-_\gamma\chi^{(\pm)+} \right)\,,\\
{\cal K}^{(\pm)-}_n=&\oint\left( 2\gamma^n\chi^{(\pm)-}\chi^{(\pm)3}-\sqrt{2}n\gamma^{n-1}\psi^-_\gamma\chi^{(\pm)-} \right)\,,
\end{aligned}
\end{equation}
where $\psi^-_\gamma$ is a shorthand for the combination $\psi^--2\gamma\psi^3+\gamma^2\psi^+$ and $\chi^{(\pm)a}$ are the $\rm S^3\times S^3$ fermions. Rewriting various fermion bilinear gives rise to the hybrid expressions \eqref{eq: spacetime R symmetry currents in hybrid variables}. The spacetime supercurrents can be constructed similarly \cite{Giveon:1998ns}, and in our convention read \cite{Gaberdiel:2024dva}
\begin{equation}
\begin{aligned}
{\cal G}^{++}_{-\frac{1}{2}}=&\oint e^{-\frac{\phi}{2}}e^{\frac{i}{2}(H_1+H_2+H_3-H_4-H_5)}\ ,\\
{\cal G}^{--}_{-\frac{1}{2}}=&\oint e^{-\frac{\phi}{2}}e^{\frac{i}{2}(H_1-H_2-H_3+H_4+H_5)}\ ,\\
{\cal G}^{-+}_{-\frac{1}{2}}=&\frac{1}{\sqrt{2}}\oint e^{-\frac{\phi}{2}}\Bigl( \, e^{\frac{i}{2}(H_1-H_2+H_3+H_4-H_5)}-\, e^{\frac{i}{2}(H_1-H_2+H_3-H_4+H_5)} \Bigr)\,,\\
{\cal G}^{+-}_{-\frac{1}{2}}=&\frac{1}{\sqrt{2}}\oint e^{-\frac{\phi}{2}}\Bigl( \, e^{\frac{i}{2}(H_1+H_2-H_3-H_4+H_5)}+\, e^{\frac{i}{2}(H_1+H_2-H_3+H_4-H_5)} \Bigr)\,.
\end{aligned}
\end{equation}
The bosons $H_i$ are the bosonisation of the 10 worldsheet fermions, see Appendix \ref{appendix: RNS to hybrid} for our convention. In the hybrid formulation, these become
\begin{equation}
\begin{split}
{\cal G}^{++}_{-\frac{1}{2}}=&\oint \Bigl( e^{-\rho}\theta^{++}c'' \Bigr)\,,\\
{\cal G}^{+-}_{-\frac{1}{2}}=&\oint \Bigl( p^{+-}-ie^{-\rho}\theta^{+-}c'' \Bigr)\,,\\
{\cal G}^{-+}_{-\frac{1}{2}}=&\oint \Bigl( p^{-+}-ie^{-\rho}\theta^{-+}c'' \Bigr)\,,\\
{\cal G}^{--}_{-\frac{1}{2}}=&\oint p^{--}\,.
\end{split}
\end{equation}
The general formula with arbitrary mode number can be obtained by considering ${\cal G}^{\alpha\beta}_{m-\frac{1}{2}}\sim[{\cal L}_m,{\cal G}^{\alpha\beta}_{-\frac{1}{2}}]$.
These DDF operators (${\cal L}_m,{\cal K}^{(\pm)a}_m,{\cal G}^{\alpha\beta}_r$), together with $({\cal U}_n,\varrho^{\alpha\beta}_r)$ that we will discuss shortly, satisfy a large ${\cal N}=4$ algebra \cite{Elitzur:1998mm}.\footnote{We have not determined the suitable normalisation for these supercurrent DDF operators which can be done by requiring the correct large ${\cal N}=4$ algebra.} 

Furthermore, it was argued \cite{Giribet:2018ada,Gaberdiel:2024dva} that the dual CFT in this tensionless limit is a free symmetric orbifold of 2 free bosons (1 compact and 1 non-compact) and 8 free fermions. Therefore, we expect to be able to write down their associated DDF operators. The DDF operator for the compact boson is easy to find following again the general procedure in \cite{Giveon:1998ns}, and the RNS result reads
\begin{equation}
\begin{split}
{\cal U}_n=\oint \left( \partial X\gamma^n+\sqrt{2}in\gamma^{n-1}\psi^-_\gamma\lambda \right)\,,
\end{split}
\end{equation}
where $X/\lambda$ is the boson/fermion in $\rm S^1$.
Applying ${\cal G}^{\alpha\beta}_{\frac{1}{2}}$, see \eqref{eq: 1/2 modes of spacetime supercurrents}, on the $-1$ mode of this DDF operator generates all its 4 superpartners $\varrho^{\alpha\beta}_{-\frac{1}{2}}$ \eqref{eq: -1/2 modes of superpartners of U_n}. Again, the formula for arbitrary mode number can be found by considering their commutators with ${\cal L}_m$.
Lastly, the DDF operator for the non-compact boson $\varphi_n$ was worked out in \cite{Sriprachyakul:2024xih}
\begin{equation}
\begin{split}
\varphi_n=&\oint\left( \gamma^n\partial\Phi+\frac{\gamma^n}{\sqrt{2}}\frac{\partial^2\gamma}{\partial\gamma}+\frac{n\gamma^{n-1}}{\sqrt{2}\partial\gamma}(\psi^--2\gamma\psi^3+\gamma^2\psi^+)(\partial\psi^--2\gamma\partial\psi^3+\gamma^2\partial\psi^+) \right)\,.
\end{split}
\end{equation}
Hence, by the same procedure, one can derive the expressions for the superpartners of $\varphi$, denoted by $\varsigma^{\alpha\beta}_r$. Therefore, we see that we can construct the DDF operators corresponding to all excitations: 2 bosonic excitations ${\cal U}_n,\varphi_n$, 8 fermionic excitations $\varrho^{\alpha\beta}_r,\varsigma^{\alpha\beta}_r$, and the large ${\cal N}=4$ excitations in the dual CFT.

\section{Conclusion and Outlook}\label{sec: conclusion and outlook}

In this article, we have discussed a Wakimoto-like free field realisation of the current algebra $\frak{d}(2,1;\alpha)_1$. This algebra plays a central role in the hybrid prescription of the tensionless string theory on $\rm AdS_3\times S^3\times S^3\times S^1$. The free field realisation discussed here requires no gauging of any additional currents as compared to \cite{Gaberdiel:2024dva}, and hence, provides a succinct, convenient description of the current algebra. Using this realisation, we explained how the Hilbert space can be constructed and showed that the full string partition function, after the appropriate on-shell projection, reproduces precisely the single-particle partition function of the dual symmetric orbifold theory. Furthermore, we discussed how various quantities in the conventional RNS formalism may be translated to the hybrid language discussed here. We believe that, given the development explored in this article, there are many interesting questions that can now be tackled more effectively. In particular, we expect to be able to perform any analogous analyses done in the tensionless limit of $\rm AdS_3\times S^3\times\mathbb{T}^4$ background. We mention a few examples of such below.

\subsection{Future directions}

\subsubsection*{Correlator computation}

It should be now straightforward, albeit possibly very tedious,\footnote{This is due to the fact that the BRST current is not as simple as in the case of tensionless strings in $\rm AdS_3\times S^3\times\mathbb{T}^4$ since $G^+$ in \eqref{eq: schematic hybrid amplitudes} is essentially the BRST current, up to some total derivative terms, see also Appendix \ref{appendix: RNS to hybrid}.} to calculate correlation functions entirely in the hybrid formulation for the background $\rm AdS_3\times S^3\times S^3\times S^1$ in the tensionless limit. In particular, the worldsheet ${\cal N}=4$ algebra \cite{Berkovits:1994vy,Berkovits:1999im} can be found by rewriting various RNS fields in terms of the hybrid fields introduced here as is briefly discussed in Appendix \ref{appendix: RNS to hybrid}. The hybrid prescription for calculating sphere correlators schematically reads \cite{Berkovits:1994vy,Berkovits:1999im}
\begin{equation}
\begin{split}
{\cal I}=\int_{{\cal M}_{0,n}}I(z_1,\cdots,z_n)\,,
\end{split}
\end{equation}
where
\begin{equation}
I(z_1,\cdots,z_n)=\left| V(z_1)(\Tilde{G}^+_0\cdot V)(z_2)(G^+_0\cdot V)(z_3)\prod_{i=4}^n(G^-_{-1}G^+_0\cdots V)(z_i) \right|^2\,.
\label{eq: schematic hybrid amplitudes}
\end{equation}
In addition to this structure, recall from \cite{Dei:2023ivl} and see also \cite{Knighton:2023mhq,Knighton:2024qxd,Sriprachyakul:2024gyl,Hikida:2023jyc}, that we have to introduce the so-called screening operator to make the holographic identification sensible. A candidate for such a screening operator, that commutes with the ${\frak d}(2,1;\alpha)_1$ currents, in our case is
\begin{equation}
{\cal O}=\int e^{\sqrt{2}\Phi}p^4\delta(\beta)\delta(\bar\beta)\,.
\label{eq: naive screening operator in the hybrid formalism}
\end{equation}
This operator is marginal, and induces a simple pole with respect to $\gamma$. Note, however, that the screening operator above is \emph{not} a rewriting of the screening operator in the RNS formulation found in \cite{Sriprachyakul:2024gyl,Knighton:2026wva}.\footnote{We should point out that it seems conceptually more natural to prefer the screening operator in \cite{Sriprachyakul:2024gyl,Knighton:2026wva} since it commutes with the ${\rm SL}(2,\mathbb{R})$ currents in the RNS formulation and it is these currents that generate the spacetime conformal symmetry \cite{Giveon:1998ns}. This is again another technical difference between our case and the tensionless strings in $\rm AdS_3\times S^3\times\mathbb{T}^4$ whose $\frak{sl}(2,\mathbb{R})$ subalgebra of $\frak{psu}(1,1|2)_1$ does generate the spacetime global conformal symmetry, see \cite{Dei:2023ivl}.
Nevertheless, it is perceivable that the operator \eqref{eq: naive screening operator in the hybrid formalism} may also give rise to holographically sensible amplitudes.}

\subsubsection*{D-branes and topological defects}
D-branes and topological defects in the tensionless string theory in $\rm AdS_3\times S^3\times\mathbb{T}^4$ were studied in \cite{Gaberdiel:2021kkp,Gutperle:2024vyp,Knighton:2024noc,Harris:2025wak}. The nice thing about these analyses is that they almost never involve the BRST operator nor the DDF operators, which are normally cumbersome in the hybrid formalism. As we have illustrated in our appendices, those operators are even more complicated in the tensionless hybrid string theory on $\rm AdS_3\times S^3\times S^3\times S^1$. Nevertheless, the partition function calculation was very simple, and thus we expect the analyses of D-branes and topological defects, which mainly involve partition functions, to be accessible. The analysis of D-branes in $\rm AdS_3\times S^3\times S^3\times S^1$ in the tensionless limit was recently done in \cite{Belleri:2025eun} in the RNS formalism. It would be useful and instructive to reproduce that result.

\subsubsection*{\boldmath $T\bar T$ deformation}

$\rm AdS/CFT$ is a staple gauge/gravity duality, it is one of the most well-studied examples. Nevertheless, we eventually want to understand more general non-AdS/non-CFT correspondences. In \cite{Dei:2024sct}, a correspondence between the tensionless strings in $\rm AdS_3\times S^3\times\mathbb{T}^4$ deformed by $J^+\bar J^+$ and a single trace $T\bar T$ deformation of the symmetric orbifold of $\mathbb{T}^4$ was explored. We believe that the ingredients that were available in that paper are now also available and hence, this analysis should be within reach.

\vspace{1cm}

Clearly, the future directions above are a few examples out of many possibilities. We hope our article provides a good reason and a good start to explore them.

\acknowledgments
I thank Matthias Gaberdiel and Bob Knighton for insightful discussions and useful comments on an early version of the draft. This work is supported by ISF grant no. 2159/22, by Simons Foundation grant 994296 (Simons Collaboration on Confinement and QCD Strings), by the Minerva foundation with funding from the Federal German Ministry for Education and Research, by the German Research Foundation through a German-Israeli Project Cooperation (DIP) grant ``Holography and the Swampland'', and by the Koshland foundation. 

\appendix

\section{\boldmath ${\frak d}(2,1;\alpha)_1$ algebra convention}\label{appendix: current algebra convention}

In this appendix, we specify our convention for ${\frak d}(2,1;\alpha)_1$ current algebra. The algebra is generated by the bosonic currents $J^a,K^{(\pm)a}$, where $a\in\{+,-,3\}$, and the fermionic currents $S^{\alpha\beta\gamma}$ where $\alpha,\beta,\gamma\in\{+,-\}$.
These currents satisfy the algebra
\begin{align}\nonumber
J^+(z)J^-(w)\sim&\frac{-2J^3(w)}{z-w}+\frac{1}{(z-w)^2}\,,\\ \nonumber
J^3(z)J^\pm(w)\sim&\pm\frac{J^\pm(w)}{z-w}\,,\\ \nonumber
J^3(z)J^3(w)\sim&-\frac{1}{2(z-w)^2}\,,\\ \nonumber
K^{(\pm)+}(z)K^{(\pm)-}(w)\sim&\frac{2K^{(\pm)3}(w)}{z-w}+\frac{2}{(z-w)^2}\,,\\ \nonumber
K^{(\pm)3}(z)K^{(\pm)\pm}(w)\sim&\pm\frac{K^{(\pm)\pm}(w)}{z-w}\,,\\ \nonumber
K^{(\pm)3}(z)K^{(\pm)3}(w)\sim&\frac{2}{2(z-w)^2}\\ \nonumber
J^a(z)S^{\alpha\beta\gamma}(w)\sim&\frac{-(\tau^a)\indices{^\alpha_\mu}S^{\mu\beta\gamma}(w)}{z-w}\,,\\ \nonumber
K^{(+)a}(z)S^{\alpha\beta\gamma}(w)\sim&\frac{-(\sigma^a)\indices{^\beta_\mu}S^{\alpha\mu\gamma}(w)}{z-w}\,,\\ \nonumber
K^{(-)a}(z)S^{\alpha\beta\gamma}(w)\sim&\frac{-(\sigma^a)\indices{^\gamma_\mu}S^{\alpha\beta\mu}(w)}{z-w}\,,\\ \nonumber
S^{\alpha\beta\gamma}(z)S^{\mu\nu\rho}(w)\sim&4(\tau_a)^{\alpha\mu}\epsilon^{\beta\nu}\epsilon^{\gamma\rho}\frac{J^a(w)}{z-w}+2(\sigma_a)^{\beta\nu}\epsilon^{\alpha\mu}\epsilon^{\gamma\rho}\frac{K^{(+)a}(w)}{z-w}\\ 
&+2(\sigma_a)^{\gamma\rho}\epsilon^{\alpha\mu}\epsilon^{\beta\nu}\frac{K^{(-)a}(w)}{z-w}-\frac{2\epsilon^{\alpha\mu}\epsilon^{\beta\nu}\epsilon^{\gamma\rho}}{(z-w)^2}\,.
\end{align}
One can massage the above algebra into the usual form of $\mathfrak{d}(2,1;\alpha)_1$ (say, the one in \cite{Gaberdiel:2024dva} with $k=1,k^\pm=2$), but we choose the current convention so that there are no additional factors of $i,\sqrt{2}$ floating around in the supercurrents \eqref{eq: fermionic d(2,1) currents}.\footnote{More precisely, we can redefine $S\to -i\sqrt{2}S$ and relate the definitions of the Pauli matrices here to the Pauli matrices in \cite{Gaberdiel:2024dva} and one would find that the OPEs look the same and we have verified this.} For completeness, we note the following identities
\begin{equation}
\begin{aligned}
(\sigma^b)\indices{^\delta_\gamma}(\sigma_b)\indices{^\alpha_\mu}=&-\frac{1}{4}\delta\indices{^\delta_\gamma}\delta\indices{^\alpha_\mu}+\frac{1}{2}\delta\indices{^\delta_\mu}\delta\indices{^\alpha_\gamma}\,,\\
(\sigma^a)\indices{_\mu^\alpha}=&-(\sigma^a)\indices{^\alpha_\mu}\,,\\
(\sigma^a)_{\alpha\beta}=&(\sigma^a)_{\beta\alpha}\,.
\end{aligned}
\end{equation}

\section{Dictionary between RNS and Hybrid formalisms}\label{appendix: RNS to hybrid}

In this appendix, we briefly summarise the process of going from the RNS to the hybrid prescriptions for tensionless strings in $\rm AdS_3\times S^3\times S^3\times S^1$, following \cite{Berkovits:1999im,Eberhardt:2019niq}. To start, we decouple the bosonic and fermionic degrees of freedom on the worldsheet, this means the matter part of the worldsheet becomes
\begin{equation}
\begin{split}
\frak{sl}(2,\mathbb{R})^{(1)}_1+\frak{su}(2)^{(1)}_2+\frak{sl}(2)^{(1)}_2+\frak{u}(1)^{(1)}=\frak{sl}(2,\mathbb{R})_3+\frak{u}(1)+\text{10 free fermions}\,.
\end{split}
\end{equation}
We denote the free fermions by $\psi^a,\chi^{(\pm)a},\lambda$, where the $\psi^a$ are the $\rm AdS_3$ fermions, $\chi^{(\pm)a}$ are the $\rm S^3$ fermions, and $\lambda$ is the $\rm S^1$ fermion. We then bosonise these fermions following \cite{Gaberdiel:2024dva}
\begin{equation}
\begin{gathered}
i\partial H_1=\psi^+\psi^-,\quad i\partial H_2=\chi^{(+)+}\chi^{(+)-},\quad i\partial H_3=\chi^{(-)+}\chi^{(-)-},\quad \\
i\partial H_4=\sqrt{2}\psi^3\left(\chi^{(+)3}+\chi^{(-)3}\right),\quad i\partial H_5=-\sqrt{2}i\lambda\left(\chi^{(+)3}-\chi^{(-)3}\right)\ ,
\end{gathered}
\end{equation}
where $H_i(z)H_j(w)\sim-\delta_{ij}\ln(z-w)$.
We also bosonise the superconformal ghosts $\hat\beta\hat\gamma$ as follows
\begin{equation}
\hat\beta=e^{-\phi+i\kappa}\partial(i\kappa),\quad \hat\gamma=e^{\phi-i\kappa}\,,
\end{equation}
where
\begin{equation}
\phi(z)\phi(w)\sim-\ln(z-w),\quad \kappa(z)\kappa(w)\sim-\ln(z-w)\,.
\end{equation}
Next, we refermionise these to
\begin{equation}
\begin{split}
p^{\alpha\beta}=&\exp\left(\frac{1}{2}(iH_1+i\alpha H_2+i\beta H_3+i\alpha\beta H_4+iH_5-\phi)\right)\,,\\
\theta^{\alpha\beta}=&\exp\left(\frac{1}{2}(-iH_1+i\alpha H_2+i\beta H_3-i\alpha\beta H_4-iH_5+\phi)\right)\,.
\end{split}
\end{equation}
and
\begin{equation}
\begin{gathered}
b'=e^{iH_1-\phi+i\kappa}=\hat{\gamma}^{-1}\psi^+,\\
c'=e^{-iH_1+\phi-i\kappa}=\hat\gamma\psi^-\,,\\
b''=e^{iH_5-\phi+i\kappa}=\hat{\gamma}^{-1}\left( \lambda-\frac{i}{\sqrt{2}}(\chi^{(-)3}-\chi^{(+)3}) \right),\\ 
c''=e^{-iH_5+\phi-i\kappa}=\hat\gamma\left( \lambda+\frac{i}{\sqrt{2}}(\chi^{(-)3}-\chi^{(+)3}) \right)\,.
\end{gathered}
\end{equation}
Lastly, the hybrid (bosonic) ghost $\rho$ is given by
\begin{equation}
\rho=2\phi-iH_1-iH_5-i\kappa\,,
\end{equation}
which has trivial OPEs with other hybrid fields above. The hybrid fields satisfy the OPEs
\begin{equation}
\begin{split}
p^{\alpha\beta}(z)\theta^{\gamma\delta}(w)\sim\frac{\epsilon^{\alpha\gamma}\epsilon^{\beta\delta}}{z-w},\quad b'(z)c'(w)\sim b''(z)c''(w)\sim\frac{1}{z-w},\quad\rho(z)\rho(w)\sim-\ln(z-w)\,.
\end{split}
\end{equation}
Additionally, one may bosonise the conformal $bc$ ghosts as
\begin{equation}
b=e^{i\sigma},\quad c=e^{-i\sigma}\,,
\end{equation}
where $\sigma(z)\sigma(w)\sim-\ln(z-w)$.

From these, one can rewrite various RNS fields in terms of the hybrid fields. For instance, one can rewrite terms in the BRST current \eqref{eq: RNS BRST current} giving
\begin{align}\nonumber
&j_{BRST}\\ \nonumber
=&-e^{2\rho+i\sigma}p^4\\ \nonumber
&+e^{2\rho}\left( p^4\left((\beta\gamma^2-3\partial\gamma-\sqrt{2}\partial\Phi\gamma)b'+\frac{i}{\sqrt{2}}\partial X b'' \right)\right.\\ \nonumber
&\hspace{1cm}\left.-\frac{i}{2}b''p^{++}(\partial p^{+-}p^{-+}-p^{+-}\partial p^{-+})p^{--} \right)\\ \nonumber
&+e^{\rho}\left( p^{++}p^{--}\left( -\beta\gamma+\frac{\partial\Phi}{\sqrt{2}}-\partial\rho-\frac{1}{2}\epsilon_{\alpha\gamma}\epsilon_{\beta\delta}p^{\alpha\beta}\theta^{\gamma\delta}-b'c' \right)+\partial p^{--}p^{++}  \right)\\ \nonumber
&+e^{\rho}\left( p^{+-}p^{-+}\left( \beta\gamma-\frac{\partial\Phi}{\sqrt{2}}+\partial\rho+\frac{1}{2}\epsilon_{\alpha\gamma}\epsilon_{\beta\delta}p^{\alpha\beta}\theta^{\gamma\delta}+b'c'+\frac{1}{2}(p^{++}\theta^{--}-p^{--}\theta^{++}) \right)\right.\\ \nonumber
&\hspace{1cm}\left.+\frac{1}{2}\partial (p^{+-}p^{-+})  \right)\\ \nonumber
&+\beta c'+\left( \frac{i}{2}(-p^{+-}\theta^{-+}+p^{-+}\theta^{+-})+\frac{i\partial X}{\sqrt{2}} \right)c''\\ \nonumber
&+e^{-i\sigma}\left( -\beta\gamma-\frac{1}{2}(\partial\Phi)^2-\frac{1}{\sqrt{2}}\partial^2\Phi-\epsilon_{\alpha\gamma}\epsilon_{\beta\delta}p^{\alpha\beta}\partial\theta^{\gamma\delta}-\frac{1}{2}(\partial X)^2-\frac{1}{2}\partial^2(3\rho+i\sigma) \right.\\ \nonumber
&\hspace{1cm}\left. -\frac{(\partial\rho)^2+(\partial i\sigma)^2}{2}+(\partial c'b'+\partial c''b'') \right)\\
&+\frac{3}{4}\partial\left( e^{-i\sigma}\left( 2\partial\rho+b'c'+b''c''+\frac{1}{2}\epsilon_{\alpha\gamma}\epsilon_{\beta\delta}p^{\alpha\beta}\theta^{\gamma\delta} \right) \right)\,,
\label{eq: the BRST current in hybrid variables}
\end{align}
where $p^4:=p^{++}p^{+-}p^{-+}p^{--}$.
Since the last term is a total derivative, one may redefine the BRST current without affecting the BRST charge by $j'_{BRST}=j_{BRST}-\tfrac{3}{4}\partial(c\hat\beta\hat\gamma)$.
Next, by rewriting various fermion bilinears, we obtain
\begin{equation}
\begin{split}
{\cal L}_m=&\oint\Biggl( \beta\gamma^{m+1}-\frac{(m+1)\gamma^m\partial\Phi}{\sqrt{2}}+\frac{m(m+1)}{2}\gamma^{m-1}e^{-\rho}(\theta^{+-}\theta^{-+}-\theta^{++}\theta^{--})c'\\
&\hspace{1cm}-(m^2-1)\gamma^m\left( \partial\rho+\frac{1}{2}\epsilon_{\alpha\gamma}\epsilon_{\beta\delta}p^{\alpha\beta}\theta^{\gamma\delta}+b'c' \right)\\
&\hspace{2cm}-\frac{m(m-1)}{2}\gamma^{m+1}e^\rho(p^{++}p^{--}-p^{+-}p^{-+})b' \Biggr)\,,
\label{eq: spacetime stress tensor in hybrid variables}
\end{split}
\end{equation}
as well as
\begin{align} \nonumber
{\cal K}^{(+)3}_n=&\oint\Biggl( \frac{1}{2}\gamma^n\left( p^{++}\theta^{--}-p^{--}\theta^{++}-p^{+-}\theta^{-+}+p^{-+}\theta^{+-} \right)\\ \nonumber
&+\frac{n\gamma^{n-1}}{2}\left( e^{-\rho}(\theta^{+-}\theta^{-+}+\theta^{++}\theta^{--})c'+i(c'b''-e^{-2\rho}\theta^4c'c'') \right)\\ \nonumber
&+\frac{n\gamma^{n+1}}{2}\left( e^{\rho}(p^{++}p^{--}+p^{+-}p^{-+})b'+i(e^{2\rho}p^4b'b''-b'c'') \right)\\ \nonumber
&+\frac{n\gamma^n}{2}\Bigl( (p^{++}\theta^{--}+p^{--}\theta^{++}+p^{+-}\theta^{-+}+p^{-+}\theta^{+-})\\ \nonumber
&\hspace{1.5cm}-ie^{\rho}(p^{++}p^{--}-p^{+-}p^{-+})+ie^{-\rho}(\theta^{+-}\theta^{-+}-\theta^{++}\theta^{--}) \Bigr) \Biggr)\,,\\ \nonumber
{\cal K}^{(+)+}_n=&\oint\Biggl( -\frac{\gamma^n}{\sqrt{2}}\Bigl( p^{++}\theta^{+-}+p^{+-}\theta^{++}-i(e^\rho p^{++}p^{+-}b''+e^{-\rho}\theta^{++}\theta^{+-}c'') \Bigr)\\ \nonumber
&-\sqrt{2}n\gamma^{n-1}e^{-\rho}\theta^{++}\theta^{+-}c'+\sqrt{2}n\gamma^{n+1}e^\rho p^{++}p^{+-}b'\\ \nonumber
&+\sqrt{2}n\gamma^n(-p^{++}\theta^{+-}+p^{+-}\theta^{++}) \Biggr)\,,\\ \nonumber
{\cal K}^{(+)-}_n=&\oint\Biggl( \frac{\gamma^n}{\sqrt{2}}\Bigl( p^{--}\theta^{-+}+p^{-+}\theta^{--}-i(e^\rho p^{-+}p^{--}b''+e^{-\rho}\theta^{-+}\theta^{--}c'') \Bigr) \\ \nonumber
&-\sqrt{2}n\gamma^{n-1}e^{-\rho}\theta^{-+}\theta^{--}c'+\sqrt{2}n\gamma^{n+1}e^\rho p^{-+}p^{--}b'\\ \nonumber
&+\sqrt{2}n\gamma^n(-p^{--}\theta^{-+}+p^{-+}\theta^{--}) \Biggr)\,,\\ \nonumber
{\cal K}^{(-)3}_n=&\oint\Biggl( \frac{1}{2}\gamma^n\left( p^{++}\theta^{--}-p^{--}\theta^{++}+p^{+-}\theta^{-+}-p^{-+}\theta^{+-} \right)\\ \nonumber
&+\frac{n\gamma^{n-1}}{2}\left( e^{-\rho}(\theta^{+-}\theta^{-+}+\theta^{++}\theta^{--})c'-i(c'b''-e^{-2\rho}\theta^4c'c'') \right)\\ \nonumber
&+\frac{n\gamma^{n+1}}{2}\left( e^{\rho}(p^{++}p^{--}+p^{+-}p^{-+})b'-i(e^{2\rho}p^4b'b''-b'c'') \right)\\ \nonumber
&+\frac{n\gamma^n}{2}\Bigl( (p^{++}\theta^{--}+p^{--}\theta^{++}+p^{+-}\theta^{-+}+p^{-+}\theta^{+-})\\ \nonumber
&+ie^{\rho}(p^{++}p^{--}-p^{+-}p^{-+})-ie^{-\rho}(\theta^{+-}\theta^{-+}-\theta^{++}\theta^{--}) \Bigr) \Biggr)\,,\\ \nonumber
{\cal K}^{(-)+}_n=&\oint\Biggl( \frac{\gamma^n}{\sqrt{2}}\Bigl( p^{++}\theta^{-+}-p^{-+}\theta^{++}-i(e^\rho p^{++}p^{-+}b''-e^{-\rho}\theta^{++}\theta^{-+}c'') \Bigr)\\ \nonumber
&+\sqrt{2}n\gamma^{n-1}e^{-\rho}\theta^{++}\theta^{-+}c'+\sqrt{2}n\gamma^{n+1}e^\rho p^{++}p^{-+}b'\\ \nonumber
&+\sqrt{2}n\gamma^n(p^{++}\theta^{-+}+p^{-+}\theta^{++}) \Biggr)\,,\\  \nonumber
{\cal K}^{(-)-}_n=&\oint\Biggl( \frac{\gamma^n}{\sqrt{2}}\Bigl( p^{--}\theta^{+-}-p^{+-}\theta^{--}-i(e^\rho p^{+-}p^{--}b''-e^{-\rho}\theta^{+-}\theta^{--}c'') \Bigr) \\ \nonumber
&-\sqrt{2}n\gamma^{n-1}e^{-\rho}\theta^{+-}\theta^{--}c'-\sqrt{2}n\gamma^{n+1}e^\rho p^{+-}p^{--}b'\\ 
&-\sqrt{2}n\gamma^n(p^{--}\theta^{+-}+p^{+-}\theta^{--}) \Biggr)\,.
\label{eq: spacetime R symmetry currents in hybrid variables}
\end{align}
As mentioned in the main text, applying ${\cal L}_m$ to ${\cal G}^{\alpha\beta}_{-\frac{1}{2}}$ generates ${\cal G}^{\alpha\beta}_{m-\frac{1}{2}}$. For instance, taking $m=1$, we obtain
\begin{equation}
\begin{split}
{\cal G}^{++}_{\frac{1}{2}}=&\oint \Bigl( e^{-2\rho}\theta^{++}\theta^{+-}\theta^{-+}c'c'' \Bigr)\,,\\
{\cal G}^{+-}_{\frac{1}{2}}=&\oint \Bigl( e^{-\rho}\theta^{+-}c'-ie^{-2\rho}\theta^{++}\theta^{+-}\theta^{--}c'c'' \Bigr)\,,\\
{\cal G}^{-+}_{\frac{1}{2}}=&\oint \Bigl( e^{-\rho}\theta^{-+}c'+ie^{-2\rho}\theta^{++}\theta^{-+}\theta^{--}c'c'' \Bigr)\,,\\
{\cal G}^{--}_{\frac{1}{2}}=&\oint \Bigl( e^{-\rho}\theta^{--}c' \Bigr)\,.
\label{eq: 1/2 modes of spacetime supercurrents}
\end{split}
\end{equation}
Applying these DDF operators to ${\cal U}_{-1}$ gives the DDF operators for its superpartners in $-\tfrac{1}{2}$ mode. Explicitly, we obtain
\begin{equation}
\begin{aligned}
\varrho^{++}_{-\frac{1}{2}}=&\oint\left( p^{++}+\gamma^{-1}e^{-\rho}\theta^{++}c' \right)\,,\\
\varrho^{+-}_{-\frac{1}{2}}=&\oint\Bigl(\left( e^{-\rho}-\gamma^{-1}e^{-2\rho}\theta^{++}\theta^{--}c' \right)\theta^{+-}c''+i\left( p^{+-}+\gamma^{-1}e^{-\rho}\theta^{+-}c' \right)\Bigr)\,,\\
\varrho^{-+}_{-\frac{1}{2}}=&\oint\Bigl(\left( e^{-\rho}-\gamma^{-1}e^{-2\rho}\theta^{++}\theta^{--}c' \right)\theta^{-+}c''+i\left( p^{-+}-\gamma^{-1}e^{-\rho}\theta^{-+}c' \right)\Bigr)\,,\\
\varrho^{--}_{-\frac{1}{2}}=&\oint\left( e^{-\rho}+\gamma^{-1}e^{-2\rho}\theta^{+-}\theta^{-+}c' \right)\theta^{--}c''\,.
\label{eq: -1/2 modes of superpartners of U_n}
\end{aligned}
\end{equation}
The DDF operator for the non-compact free boson reads
\begin{equation}
\begin{split}
\varphi_n=&\oint\Biggl( \gamma^n\partial\Phi+\frac{\gamma^n}{\sqrt{2}}\frac{\partial^2\gamma}{\partial\gamma}+\frac{n\gamma^{n-1}}{\sqrt{2}\partial\gamma}e^{-2\rho}\theta^4c'\partial c'+\frac{n\gamma^{n+3}}{\sqrt{2}\partial\gamma}e^{2\rho}p^4b'\partial b'\\
&-\frac{n\gamma^n}{\sqrt{2}\partial\gamma}\left( \partial\left( e^{-\rho}c' \right)-e^{-\rho}c'\epsilon_{\alpha\gamma}\epsilon_{\beta\delta}p^{\alpha\beta}\theta^{\gamma\delta} \right)(\theta^{+-}\theta^{-+}-\theta^{++}\theta^{--})\\
&-\frac{n\gamma^{n+2}}{\sqrt{2}\partial\gamma}\left( \partial\left( e^{\rho}b' \right)+e^{\rho}b'\epsilon_{\alpha\gamma}\epsilon_{\beta\delta}p^{\alpha\beta}\theta^{\gamma\delta} \right)(p^{++}p^{--}-p^{+-}p^{-+}) \\
&+\frac{n\gamma^{n+1}}{\sqrt{2}\partial\gamma}\left( c'\partial b'+b'\partial c' -\partial^2\rho-\frac{1}{4}\epsilon_{\alpha\gamma}\epsilon_{\beta\delta}\partial p^{\alpha\beta}\theta^{\gamma\delta}-\frac{3}{4}\epsilon_{\alpha\gamma}\epsilon_{\beta\delta}p^{\alpha\beta}\partial\theta^{\gamma\delta}\right.\\
&\hspace{2cm}-(p^{++}p^{--}-p^{+-}p^{-+})(\theta^{+-}\theta^{-+}-\theta^{++}\theta^{--}) \\
&\hspace{2.5cm}\left.-2b'c'\left( \partial\rho+\frac{1}{2}\epsilon_{\alpha\gamma}\epsilon_{\beta\delta}p^{\alpha\beta}\theta^{\gamma\delta} \right)-\frac{1}{4}\left( \epsilon_{\alpha\gamma}\epsilon_{\beta\delta}p^{\alpha\beta}\theta^{\gamma\delta} \right)^2 \right) \Biggr)\,.
\end{split}
\end{equation}
Its superpartners can be found similarly to what was done for the spacetime $U(1)$, however, since the expressions are cumbersome and we do not need them currently, we will not write them down.

In addition, one can also construct the worldsheet (small) ${\cal N}=4$ algebra following \cite{Berkovits:1999im}. To start, equation (3.1) of \cite{Berkovits:1999im} in our convention reads
\begin{equation}
\begin{split}
T=&-\beta\gamma-\frac{1}{2}(\partial\Phi)^2-\frac{Q}{2}\partial^2\Phi-\frac{1}{2}(\partial X)^2-\epsilon_{\alpha\gamma}\epsilon_{\beta\delta}p^{\alpha\beta}\partial\theta^{\gamma\delta}\\
&+\frac{(\partial i\sigma)^2}{2}-\partial^2(i\sigma)-\frac{(\partial\rho)^2}{2}-\partial^2\rho+\frac{\partial b'c'+\partial c'b'+\partial b''c''+\partial c''b''}{2}\,,\\
G^+=&j_{BRST}-\frac{3}{4}\partial\left(e^{-i\sigma}\left(2\partial\rho+b'c'+b''c''+\frac{1}{2}\epsilon_{\alpha\gamma}\epsilon_{\beta\delta}p^{\alpha\beta}\theta^{\gamma\delta} \right)\right)+\partial^2e^{-i\sigma}\\
&+\partial(e^{-i\sigma}(\partial\rho+b'c'+b''c''))\,,\\
G^-=&e^{i\sigma}\,,\\
J=&-(\partial\rho+i\partial\sigma+b'c'+b''c'')\,.
\label{eq: N=2 generators in hybrid}
\end{split}
\end{equation}
These fields generate the standard ${\cal N}=2$ algebra with central charge $c=6$
\begin{equation}
\begin{split}
T(z)T(w)\sim&\frac{6/2}{(z-w)^4}+\frac{2T(w)}{(z-w)^2}+\frac{\partial_wT}{z-w}\,,\\
T(z)G^\pm(w)\sim&\frac{3}{2}\frac{G^\pm(w)}{(z-w)^2}+\frac{\partial_wG^\pm}{z-w}\,,\\
T(z)J(w)\sim&\frac{J(w)}{(z-w)^2}+\frac{\partial_wJ}{z-w}\,,\\
G^+(z)G^-(w)\sim&\frac{6/3}{(z-w)^3}+\frac{J(w)}{(z-w)^2}+\frac{T(w)+\tfrac{1}{2}\partial_wJ}{z-w}\,,\\
J(z)J(w)\sim&\frac{6/3}{(z-w)^2}\,,\\
J(z)G^\pm(w)\sim&\pm\frac{G^\pm}{z-w}\,,\\
G^-(z)G^-(w)\sim G^+(z)G^+(w)\sim&0\,.
\end{split}
\end{equation}
Together with the additional fields 
\begin{equation}
\begin{split}
J^{++}=&-e^{-\rho-i\sigma}c'c''\,,\\
J^{--}=&-e^{\rho+i\sigma}b'b''\,,\\
\Tilde{G}^+=&e^{-\rho}c'c''\,,\\
\Tilde{G}^-=&(J^{--}\cdot G^+)=\frac{1}{2\pi i}\oint_zdwJ^{--}(w)G^+(z)\,,\\
\end{split}
\end{equation}
they generate an (untwisted, small) ${\cal N}=4$ algebra. Following Section 2 of \cite{Berkovits:1999im}, one may perform the topological twist which amounts to adding $\partial J/2$ to the stress tensor $T$, that is, $T^{\text{top. twist}}=T+\partial J/2$. This then allows us to rewrite the RNS physical state conditions as
\begin{equation}
T^{\text{top. twist}}_0V=G^+_0V=\Tilde{G}^+_0V=(J_0-1)V=0,\quad V\sim V+G^+_0\Tilde{G}^+_0\Lambda\,.
\end{equation}
The first condition simply says that the physical vertex operators have vanishing conformal weight.\footnote{For instance, in the RNS formalism, the gauge-fixed, 0-picture physical vertex operators have the form $cV^{\rm matter}_{\Delta=1}$ which have vanishing total conformal weight. This property guarantees that the vertex operators do not transform under the conformal transformation on the worldsheet.} Note also that $T^{\text{top. twist}}$ is just the RNS, ${\cal N}=1$ matter$+$ghost stress tensor. The $G^+_0$ condition is the usual BRST condition and the $\Tilde{G}^+_0$ condition is the condition that vertex operators do not depend on the zero mode of $\xi$ where $\xi$ is the bosonisation of the superconformal ghosts $\hat\beta=e^{-\phi}\partial\xi,\hat\gamma=e^\phi\eta$. The latter condition is necessary to ensure that physical vertex operators are constructible in terms of the unbosonised $\hat\beta,\hat\gamma$ ghosts. The charge $J_0$ constraint is the familiar relation between the ghost and picture numbers and the cohomology equivalence is the BRST cohomology equivalence, see also footnote 5 of \cite{Berkovits:1999im}.

\section{Theta function convention}\label{appendix: theta function convention}

In this appendix, we define our convention for the Jacobi theta functions and the Dedekind eta function. These functions are defined as follows
\begin{equation}
\begin{split}
\eta(\tau)=q^{\frac{1}{24}}\prod_{n=1}^\infty(1-q^n)\,,
\end{split}
\end{equation}
whereas
\begin{equation}
\begin{split}
\vartheta_1(z;\tau)=&-iq^{\frac{1}{8}}(y^{\frac{1}{2}}-y^{-\frac{1}{2}})\prod_{n=1}^\infty(1-q^n)(1-yq^n)(1-y^{-1}q^n)\\
=&-i\sum_{n=-\infty}^\infty (-)^ny^{n+\frac{1}{2}}q^{\frac{1}{2}(n+\frac{1}{2})^2}\,,\\
\vartheta_2(z;\tau)=&q^{\frac{1}{8}}(y^{\frac{1}{2}}+y^{-\frac{1}{2}})\prod_{n=1}^\infty(1-q^n)(1+yq^n)(1+y^{-1}q^n)\\
=&\sum_{n=-\infty}^\infty y^{n+\frac{1}{2}}q^{\frac{1}{2}(n+\frac{1}{2})^2}\,,\\
\vartheta_3(z;\tau)=&\prod_{n=1}^\infty(1-q^n)(1+yq^{\frac{2n-1}{2}})(1+y^{-1}q^{\frac{2n-1}{2}})\\
=&\sum_{n=-\infty}^\infty y^{n}q^{\frac{1}{2}n^2}\,,\\
\vartheta_4(z;\tau)=&\prod_{n=1}^\infty(1-q^n)(1-yq^{\frac{2n-1}{2}})(1-y^{-1}q^{\frac{2n-1}{2}})\\
=&\sum_{n=-\infty}^\infty (-)^ny^{n}q^{\frac{1}{2}n^2}\,.
\end{split}
\end{equation}

\bibliography{bibliography}
\bibliographystyle{utphys.sty}

\end{document}